\pdfoutput=1
\documentclass[copyright,creativecommons]{eptcs}
\providecommand{\event}{QPL 2022} %

\ifpdf
  \usepackage{underscore}         %
  \usepackage[T1]{fontenc}        %
\else
  \usepackage{breakurl}           %
\fi
\usepackage[utf8]{inputenc}
\usepackage[english]{babel}
\usepackage{csquotes}
\usepackage[maxbibnames=99,sorting=nyt,style=numeric]{biblatex}
\usepackage{amsmath,amssymb,amsthm,mathtools}
\usepackage{proof,mathpartir}
\usepackage{colonequals}
\usepackage{booktabs}
\usepackage{bm}

\usepackage{pfpl-syntax}

\usepackage{braket}
\usepackage[usenames,dvipsnames]{xcolor}
\usepackage{microtype}

\usepackage{xspace}
\DeclareMathAlphabet{\pazocal}{OMS}{zplm}{m}{n}
\newcommand{\qwire}{$\pazocal{Q}$\textsc{wire}\xspace}

\newcommand{\qsharp}{Q\#\xspace}
\newcommand{\core}{\ensuremath{\lambda_{\qsharp}}\xspace}

\newcommand{\algol}{A\textsc{lgol}\xspace}
\newcommand{\ALGOL}{A{\small LGOL}\xspace}

\ProvideDocumentCommand{\PFPL}{}{\textbf{\textsf{PFPL}}\xspace}
\ProvideDocumentCommand{\MAPFPL}{}{\textbf{\textsf{MA}}\xspace}
\ProvideDocumentCommand{\eqdef}{}{\mathrel{\triangleq}}

\newcommand{\bnfdef}{\coloncolonequals}
\newcommand{\bnfalt}{\mathrel{\blk{\mid}}}

\newcommand{\Sort}[1]{\textsf{#1}}

{\begin{equation*}
      \renewcommand{\bnfalt}{\mathrel{\phantom{\mid}}}
      \begin{array}{llc@{\quad\extracolsep{\fill}}lll}
}
{\end{array}\end{equation*}\ignorespacesafterend}

\newenvironment{synchart*}[1]%
{\begin{equation*}
      \renewcommand{\bnfalt}{\mathrel{\phantom{\mid}}}
      \begin{array}{llc@{\quad\extracolsep{\fill}}lll}
        \textit{Sort} & & & \textit{Abstract} & \textit{Concrete} & \\
}
{\end{array}\end{equation*}\ignorespacesafterend}

\NewDocumentCommand{\letCst}{m m m m}{\Opn{let}\,\blk{#3}\,\mathbin{\kw{be}}\,#2\,\mathbin{\kw{in}}\,#4}
\NewDocumentCommand{\letAbt}{m m m m m}{\Abt{\Opn{let}}[#1\sep #2]{#3\sep \Abs(#4){#5}}}
\NewDocumentCommand{\letAbtSugar}{m m m}{\Abt{\Opn{let}}{#1\sep \Abs(#2){#3}}}
\NewDocumentCommand{\letbnd}{s O{\tau_1} m O{\tau_2} m m}{\IfBooleanTF{#1}{\letCst{#4}{#3}{#5}{#6}}{\letAbt{#2}{#4}{#3}{#5}{#6}}}

\NewDocumentCommand{\qrefCst}{m}{\texttt{\&}\,\qbit{#1}}
\NewDocumentCommand{\qrefAbt}{m}{\Abt{\Opn{qref}}<\qbit{#1}>{}}
\NewDocumentCommand{\qrefty}{s m}{\IfBooleanTF{#1}{\qrefCst{#2}}{\qrefAbt{#2}}}

\NewDocumentCommand{\newCst}{m m m}{{\Opn{new}}~\blk{#2}\,\mathbin{\kw{in}}\,#3}
\NewDocumentCommand{\newAbt}{m m}{\Abt{\Opn{newqref}}[#1]{#2}}
\NewDocumentCommand{\newAbtSugar}{m m}{\Abt{\Opn{newqref}}{\Abs<#1>{#2}}}
\NewDocumentCommand{\newcmd}{s O{\tau} m m}{\IfBooleanTF{#1}{\newCst{#2}{#3}{#4}}{\newAbt{#2}{\Abs<#3>{#4}}}}

\NewDocumentCommand{\gateCst}{m m}{\Abt{\Opn{apply}\nolimits_{\igate{#1}}}{#2}}
\NewDocumentCommand{\gateAbt}{m m m}{\Abt{\Opn{gateap}}<\igate{#2_{#1}}>{#3}}
\NewDocumentCommand{\gateap}{s O{2^n} m m}{\IfBooleanTF{#1}{\gateCst{#3}{#4}}{\gateAbt{\blk{#2}}{#3}{#4}}}

\NewDocumentCommand{\diagCst}{m m m m}{\Abt{\Opn{apply}\nolimits_{D(\igate{#1},\igate{#2})}}{#3\sep #4}}
\NewDocumentCommand{\diagAbt}{m m m m m}{\Abt{\Opn{diagap}}<\igate{#2_{#1}},\igate{#3_{#1}}>{#4\sep #5}}
\NewDocumentCommand{\diagap}{s O{2^n} m m m m}{\IfBooleanTF{#1}{\diagCst{#3}{#4}{#5}{#6}}{\diagAbt{\blk{#2}}{#3}{#4}{#5}{#6}}}

\NewDocumentCommand{\measCst}{m}{\Abt{\Opn{meas}}{#1}}
\NewDocumentCommand{\measAbt}{m}{\Abt{\Opn{meas}}{#1}}
\NewDocumentCommand{\meascmd}{s m}{\IfBooleanTF{#1}{\measCst{#2}}{\measAbt{#2}}}

\NewDocumentCommand{\bndSugar}{m m m}{\blk{#2}\mathbin{\leftarrow}#1\mathbin{\kw{;}}#3}
\NewDocumentCommand{\ifSugar}{m m m}{\Opn{if}\,#1\,\Opn{then}\,#2\,\Opn{else}\,#3}

\newcommand{\mailtodomain}[1]{\href{mailto:#1}{\nolinkurl{#1}}}

\providecolor{ACMPurple}{RGB}{101,1,107}
\providecolor{ACMDarkBlue}{RGB}{9,53,122}

\usepackage{hyperxmp}
\def\titlerunning{\qsharp as a Quantum Algorithmic Language}
\hypersetup{
    colorlinks,
    linktoc=all,
    bookmarksnumbered,
    linkcolor=ACMPurple,
    citecolor=ACMPurple,
    urlcolor=ACMDarkBlue,
    filecolor=ACMDarkBlue,
    keeppdfinfo,
    pdftitle={\titlerunning},
    pdfauthor={Kartik Singhal, Kesha Hietala, Sarah Marshall, and Robert Rand},
    pdfsubject={The Essence of \qsharp},
    pdfkeywords={\qsharp, quantum computation, type systems, formal specification, formal language definitions, semantics and reasoning, quantum programming languages, language design},
    pdfcontactemail={ks@cs.uchicago.edu},
    pdfcontacturl={https://ks.cs.uchicago.edu},
    pdfdate={2023-03-16},
    pdfidentifier={arXiv:2206.03532},
    pdflicenseurl={https://creativecommons.org/licenses/by/4.0/},
    pdfpubstatus={VoR},
    pdfpubtype={report},
    pdfurl={https://ks.cs.uchicago.edu/publication/q-algol/q-algol.pdf},
    pdfversionid={v3},
    pdfapart=2,
    pdfaconformance=B
}
\immediate\pdfobj stream attr{/N 3} file{sRGB.icc}
\pdfcatalog{%
  /OutputIntents [
    <<
      /Type /OutputIntent
      /S /GTS_PDFA1
      /DestOutputProfile \the\pdflastobj\space 0 R
      /OutputConditionIdentifier (sRGB)
      /Info (sRGB)
    >>
  ]
}

\usepackage[figure,table]{hypcap}

\title{\titlerunning}

\author{
\href{https://orcid.org/0000-0003-1132-269X}{Kartik Singhal}
\institute{University of Chicago}
\email{\mailtodomain{ks@cs.uchicago.edu}}
\and
\href{https://orcid.org/0000-0002-2724-0974}{Kesha Hietala}
\institute{University of Maryland}
\email{\mailtodomain{kesha@cs.umd.edu}}
\and
\href{https://orcid.org/0000-0002-7744-4932}{Sarah Marshall}
\institute{Microsoft Quantum}
\email{\mailtodomain{sarah@sarahmarshall.name}}
\and
\href{https://orcid.org/0000-0001-6842-5505}{Robert Rand}
\institute{University of Chicago}
\email{\mailtodomain{rand@uchicago.edu}}
}

\def\authorrunning{K. Singhal, K. Hietala, S. Marshall, and R. Rand}

\usepackage{subcaption}
\usepackage{listings}
\usepackage{qsharp}
\lstdefinelanguage{lqsharp}{
    language     = qsharp
}
\usepackage[capitalize,nameinlink,noabbrev]{cleveref}
\crefformat{section}{#2\S#1#3}
\crefformat{lstlisting}{#2Listing~#1#3}

\usepackage{ottalt}

\newenvironment{ottdefnblock}[3][]{ \framebox{\mbox{#2}} \quad #3 \\[0pt]}{}

\newcommand{\ottafterlastrule}{\\}
  \usepackage[mathx]{mathabx}

  \usepackage{stackengine}
  \usepackage{scalerel}
  \stackMath
  \newcommand{\ccdot}{\scalebox{1.15}{$\SavedStyle\cdot$} }
  \newcommand{\altdiv}{\mathbin{\ThisStyle{%
    \stackunder[-.25\LMex]{\stackon[-.15\LMex]{\SavedStyle\sim}{\ccdot} }{\ccdot} } } }

  \renewottcommands[ott]

\usepackage{tabularx}
\newcolumntype{L}{>{$}l<{$}} %

\newcommand{\s}[1]{\mathrm{#1~}}
\newcommand{\snsp}[1]{\mathrm{#1}} %
\newcommand{\su}[1]{\underline{\mathsf{#1}}}

\newcommand{\blk}[1]{{\textcolor{black}{#1}}}
\newcommand{\blue}[1]{{\textcolor{RoyalBlue}{#1}}}
\newcommand{\qbit}[1]{{\textcolor{RedOrange}{#1}}}
\newcommand{\igate}[1]{\ensuremath{\textcolor{CarnationPink}{\mathit{#1}}}}

\newcommand{\docmd}[1]{\s{\kw{do}}\{#1\}}

\newcommand{\lqs}[1]{\lstinline{#1}\xspace}
\newcommand{\lqsnsp}[1]{\lstinline{#1}} %

\newcommand{\elabl}{[\![} %
\newcommand{\elabr}{]\!]}
\newcommand{\elab}[1]{\blk{\elabl{#1}\elabr}}

\theoremstyle{plain}
\newtheorem{theorem}{Theorem}
\newtheorem{proposition}{Proposition}

\theoremstyle{remark}

\theoremstyle{definition}
\newtheorem{example}{Example}[section]

\begin{document}
\maketitle

\begin{abstract}
    \qsharp is a standalone domain-specific programming language from Microsoft for writing and running quantum programs.
    Like most industrial languages, it was designed without a formal specification, which can naturally lead to ambiguity in its interpretation.
    We aim to provide a formal language definition for \qsharp, placing the language on a solid mathematical foundation and enabling further evolution of its design and type system.
    This paper presents \core, an idealized version of \qsharp that illustrates how we may view \qsharp as a quantum \algol (algorithmic language).
    We show the safety properties enforced by \core's type system
    and present its equational semantics based on a fully complete algebraic theory by Staton.
\end{abstract}

\section{Introduction}
\label{sec:intro}

Microsoft's \qsharp programming language~\autocite{Svore2018} is one of the most full-featured quantum programming languages that have emerged from the recent boom in quantum computing research.
However, with a growing code base and increasing popularity comes the demand for more features and the resulting added complexity. Hence, \qsharp faces challenges familiar to many growing programming languages---maintaining correctness, ease of use, and intuitive understanding while evolving to meet users' needs.

Quantum programming languages face unique challenges that are not present in classical languages.
Quantum algorithms are more challenging to design and reason about than classical ones as they use quantum phenomena like superposition and entanglement.
Quantum programs are challenging to test and debug.
Their simulation on classical computers is slow and limited to a handful of qubits while languages like \qsharp are designed for large-scale fault-tolerant quantum computers with thousands of logical qubits.
When running a quantum program directly on quantum hardware, we cannot observe the whole quantum state directly, and measuring a classical result during execution (partial observation) may itself destroy the state.
Additionally, existing quantum hardware provides limited qubit count and poor gate fidelity.

These challenges underscore why it is essential for \qsharp to have a well-specified definition that can serve as a foundation for extensions, multiple implementations, and formal verification of programs written in the language. A formal specification and mechanization of its metatheory will help ensure that \qsharp is robust enough to meet the unique needs of the developing field of quantum software engineering.

A tried-and-tested approach to achieving this ambitious goal is to define an idealized core version of the language, provide an elaboration from the surface language to this core language, and provide static and dynamic semantics for the core. In this paper, we argue that even though \qsharp is a relatively large language, we can condense it to a small core capturing most of its interesting features.
We call this core \core.
In it, we make several implicit features of \qsharp explicit: its treatment of qubits as references, its stack-like memory management that enables reasoning about the quantum state in a local manner, and its safe synthesis of effectful and pure computation.
In the classical setting, this stack-like memory management and combination of effectful and pure computation are inherent to many \algol-like languages~\autocite{Reynolds1997}.

\paragraph{Contributions}
In 2018, when introducing \qsharp, its designers stated that as opposed to several existing \textit{circuit definition} languages, ``\qsharp is an \textit{algorithm definition} language''~\autocite{Svore2018}; the goal of our paper is to show that, in its essence, \qsharp is a quantum \textit{algorithmic language} (\algol).
\label{sec:contrib}
\begin{itemize}
  \item To support this characterization, we introduce \core, an idealized version of \qsharp inspired by Harper's language \MAPFPL (Modernized Algol)~\autocite{Harper2016a}.
  In \core, we expose values of the \lqs{Qubit} type in \qsharp as references to logical qubits and formalize the \algol-like stack discipline implicit in \qsharp's quantum memory management.
  \item We develop a type system for \core that extends \qsharp's type system to enforce the no-cloning theorem and stack-like management of qubits. %
  \item We provide an equational dynamics for \core building upon the fully complete equational theory of quantum computation by Staton~\autocite{Staton2015}.
  \item Finally, we provide an elaboration relation from \qsharp to \core, thereby endowing a significant portion of \qsharp with a formal specification and additional safety guarantees.
\end{itemize}

\paragraph{Outline}
\label{sec:outline}
In the rest of the paper, we review background on the \qsharp programming language and Staton's theory for quantum computation (\cref{sec:background}); introduce \core along with its syntax and semantics (\cref{sec:lambdaqs}); describe how \core is faithful to the surface \qsharp language (\cref{sec:elaboration}); and discuss related and future work (\cref{sec:relatedwork} and \cref{sec:conclusion}).

\section{Background}
\label{sec:background}

Before introducing \core, we discuss the two projects that inspired our work. The first is Microsoft's \qsharp~\autocite{Svore2018}, a modern, self-contained quantum programming language that
boasts a large community of developers. The second is Sam Staton's equational theory for quantum programs~\autocite{Staton2015}, which provides a compelling alternative to the standard matrix-based semantics for quantum programs.

\subsection{The \qsharp Programming Language}
\label{sec:qsharp}

\qsharp~\autocite{Svore2018} is a hybrid quantum-classical programming language that supports interlacing stateful quantum \emph{operations} with pure classical \emph{functions}, collectively referred to as \emph{callables}.
\qsharp encourages thinking about quantum programs as algorithms rather than circuits, where quantum operations can be combined with classical control flow such as branches and loops.
When a programmer measures a qubit, they can perform an arbitrary classical computation on the result, and the program execution can continue without requiring the qubit to be released.
This computational model allows quantum and classical algorithms to be fully mixed.
At the same time, \qsharp enforces a degree of separation between the quantum and classical components.
Operations can call functions, but functions cannot call operations.
An example \qsharp program, implementing the quantum teleportation protocol, is shown in \Cref{lst:teleport} in \Cref{app:elaboration}.

\qsharp contains a blend of functional and imperative features (it evolved from an F\#-like language~\autocite{Azariah2018}).
For classical data, \qsharp follows the so-called value semantics~\autocite{Heim2020a}, perhaps better known as referential transparency. \qsharp functions are always pure, and variable bindings are immutable by default. Bindings may be declared \lqs{mutable}, but they correspond to a local state change, enclosed in the scope of the parent callable. Hence, equational reasoning is possible across function boundaries.
By contrast, qubits are opaque types that act as references to logical qubits~\autocite{Geller2018, Geller2019}---their values are never exposed.
Gate operations are inherently \emph{effectful}: a (single-qubit) quantum gate application is a procedure that takes a qubit reference as input and returns a trivial output of type \lqs{Unit} after altering the quantum state.

Callables in \qsharp can be \emph{higher-order}: functions and operations are values and can be given as arguments to, or returned by, other functions and operations. Both functions and operations can be partially applied. Quantum algorithms parameterized by quantum subroutines are easily expressed in \qsharp using higher-order operations. For example, an operation implementing Grover's search~\autocite{Grover1996} can accept an oracle as a parameter and apply it in each iteration.

\qsharp supports a restricted form of metaprogramming, where the compiler can automatically generate the adjoint and controlled versions of unitary operations.
Operations can declare their support for \lqs{Adjoint} and \lqs{Controlled} \emph{functors} (in \qsharp's terminology\footnote{Perhaps a better name for functors would be `combinators' from the functional programming community to avoid confusion with other accepted meanings of the term `functor.'}) using the \emph{characteristics} \lqs{Adj} and \lqs{Ctl}, respectively.

\qsharp follows the QRAM model of computation~\autocite{Knill1996}, which assumes an unbounded supply of logical qubits from which the programmer can obtain a reference to a new qubit by calling the \lqs{use} command. Qubits are hence allocated and deallocated in a stack-like manner, where the lifetime of a qubit is equivalent to the lexical scope of the \lqs{use} command. Even though this stack discipline can ensure safe (quantum) memory management, it is currently not enforced by the \qsharp compiler and type system. \Cref{lst:stack} shows a minimal example that passes the type checker but fails at runtime (in a simulator).

\begin{figure}
  \begin{subfigure}{0.48\textwidth}
    \begin{lstlisting}
    operation NewQubit () : Qubit {
      use q = Qubit();
      return q;
    }
    \end{lstlisting}
  \caption{Returning a qubit after its lifetime has ended.}
  \label{lst:stack}
  \end{subfigure}
  \hfill
  \begin{subfigure}{0.48\textwidth}
    \begin{lstlisting}
    operation Clone () : Unit {
      use q1 = Qubit();
      let q2 = q1;
      CNOT(q1, q2);
    }
    \end{lstlisting}
  \caption{Using the same qubit as both control and target.}
  \label{lst:alias}
  \end{subfigure}
  \caption{Sample unsafe \qsharp programs.}
\end{figure}

Programmers are allowed to create new bindings using \lqs{let} that refer to the same qubit as another binding, leading to \emph{aliasing} of qubit references.
While aliasing is ubiquitous in \qsharp, it can lead to unsafe behavior in violation of the \emph{no-cloning theorem}~\autocite{Wootters1982}, which forbids duplication of qubits. In \Cref{lst:alias}, both \lqs{q1} and \lqs{q2} refer to the same qubit. Applying \lqs{CNOT} with \lqs{q1} as the control and \lqs{q2} as the target is equivalent to cloning the underlying qubit. Currently, \qsharp cannot prevent this issue statically.

An informal specification of the \qsharp language was recently published~\autocite{QsSpec2020}. However, it does not capture the subtle aspects of the language, such as the aliasing of qubit references or its goal of maintaining a stack discipline. Our work makes these subtleties explicit and formal.

\subsection{An Equational Theory for QRAM}
\label{sec:qramtheory}

Staton~\autocite{Staton2015} presents a substructural (linear) version of his framework for ``parameterized algebraic theories''~\autocite{Staton2013}. He develops an axiomatization for quantum computation using this framework, which he shows to be fully complete. Staton then extracts an equational theory for a quantum programming language from his algebraic theory that uses \textit{generic effects} rather than \textit{algebraic operations}~\autocite{Plotkin2003}.
Finally, Staton remarks upon a variant of his theory~\autocite[\S 6.2]{Staton2015} that applies to the QRAM model, where instead of working with qubits, we work with references to qubits. This is the approach taken in projects like the Quantum IO Monad~\autocite{Altenkirch2009}, Quantum Hoare Type Theory~\autocite{Singhal2020, Singhal2021}, and, to our advantage, \qsharp.\footnote{However, the stack-like management of qubits is unique to \qsharp.}

Here we reproduce Staton's theory of a ``quantum local store''~\autocite[\S 6.2, p. 11]{Staton2015} for reference; we will see in \cref{sec:dynamics} how this algebraic theory helps us describe the equational dynamics of \core.
We assume that the qubit references are unique, which we guarantee for \core in \cref{para:safety}.

\paragraph{Generic Effects}
Staton adds the following generic effects to a standard linear type theory and obtains a quantum programming language~\autocite[\S 5]{Staton2015} similar to Selinger's QPL~\autocite{Selinger2004}.
\begin{mathpar}
  \inferrule{ }{\vdash \su{new}() : \s{qubit}} \and
  \inferrule{\Gamma \vdash t : \snsp{qubit}^{\otimes n}}{\Gamma \vdash \su{apply}_U(t) : \snsp{qubit}^{\otimes n}}  \and
  \inferrule{\Gamma \vdash t : \s{qubit}}{\Gamma \vdash \su{measure}(t) : \s{bool}}
\end{mathpar}

\paragraph{Program Equations}
There are two interesting classes of axioms (ignoring the axioms that describe commutativity of $\snsp{let}$). For completeness, we also show axiom (C) pertaining to the \textit{discard} operation (equivalent to measuring a qubit and ignoring its result). However, it does not apply in the QRAM model as noted by Staton~\autocite[\S 6.2]{Staton2015}.

\noindent Axioms relating unitary gates and measurement:
\begin{align*}
  (A) \quad && \su{measure}(\su{apply}_X(a)) &\equiv{} \neg~\su{measure}(a) \\
  (B) \quad && \s{let}(a', x')~\s{be} \su{apply}_{D(U,V)}(a, x)&~\s{in}(\su{measure}(a'), x') \equiv{} \\
   && \s{if}\su{measure}(a)&=0~\s{then}(0, \su{apply}_U(x))~\s{else}(1, \su{apply}_V(x)) \\
  (C) \quad && \su{discard}(\su{apply}_U(x)) &\equiv{} \su{discard}(x)
\end{align*}

\noindent Axioms relating allocation with unitaries and measurement:
\begin{alignat*}{3}
  (D) \quad &&
  \qquad\qquad\qquad \su{measure}(\su{new}()) &\equiv 0 \\
  (E) \quad &&
  \qquad\qquad\qquad \su{apply}_{D(U,V)}(\su{new}(), x) &\equiv (\su{new}(), \su{apply}_U(x))
\end{alignat*}
\noindent where $D(U, V) = U \oplus V = \big(\begin{smallmatrix}U & 0 \\ 0 & V\end{smallmatrix}\big)$
applies $U$ or $V$ depending on the value of its first argument.

Axiom (A) says that applying the quantum X gate to a qubit and then measuring it is the same as negating the measurement result. Axiom (B) explains the action of a block diagonal matrix $D(U, V)$ as quantum control by stating that applying the diagonal matrix and then measuring the control qubit is equivalent to measuring the control qubit and branching on the result to decide whether to apply $U$ or $V$. Axiom (C) says that if the qubits are to be discarded, applying a unitary is the same as doing nothing. Axiom (D) states that measuring a new qubit always results in $0$, i.e., qubits are always initialized to $0$. Axiom (E) says that using a new qubit as control is the same as controlling by $0$.

We will show in \cref{sec:dynamics} that our \core calculus follows similar program equations.

\section{\texorpdfstring{$\bm{\lambda}_{\mathbf{\qsharp}}$}{λ-\qsharp}: A Core Calculus for \qsharp}
\label{sec:lambdaqs}

Our approach closely follows the type-theoretic interpretation of Standard ML, where Harper and Stone~\autocite{Harper2000} developed a well-typed internal language for Standard ML, defined an elaboration relation between the external language and this internal language, and proved the properties of the metatheory of the language using the internal language.
Harper and collaborators~\autocite{Crary2009, Lee2007} followed this work with the mechanization of the metatheory using the Twelf logical framework~\autocite{Pfenning1999}.
As a first step, we identify and isolate the core language, \core, that captures the essential aspects of \qsharp. This core language is explicitly typed, and the safety properties of its type structure can be easily stated and proved.

Once we have identified the core, we define an elaboration relation from the surface \qsharp language to \core (\cref{sec:elaboration}). A \qsharp program is well-formed when it has a well-typed elaboration, and its semantics is defined to be that of its elaboration. The advantage of this approach is that proving properties about the metatheory of a large language becomes tractable because we only need to do it for the small, well-formed core.

To mirror the separation between operations and functions in \qsharp, we base the design of \core on Harper's \MAPFPL (Modernized Algol)~\autocite{Harper2016a}, which maintains a separation between commands that modify state and expressions that do not.
\qsharp is an \algol-like language in more ways than one---syntax, block-structure, local (classical) state, and safe integration of functional and imperative paradigms.
However, unlike Reynolds' Idealized \algol~\autocite{Reynolds1997},
the variables in \qsharp are immutable by default, and the language follows a call-by-value semantics, both of which make it closer to Harper's \MAPFPL.

Before presenting \core, let us motivate our design choices and establish some terminology.
\qsharp has two kinds of variable bindings.
Those defined using the \lqs{let} keyword are the same as the variables in \MAPFPL and follow the usual substitution-based semantics of functional programming languages.
Those defined using the \lqs{mutable} keyword correspond to \textit{assignables} that can be reassigned similar to ``variables'' in imperative languages. Significantly, they are restricted to the lexical scope in which they are bound.
Since they do not affect equational reasoning across function boundaries, and our focus is on the quantum state, we ignore \lqs{mutable} variables in the rest of this paper.
Qubits have type \lqs{Qubit} and syntactically look just like other variables but are references to underlying logical qubits that are never exposed.
Unlike classical bindings, which follow value semantics, aliasing is permitted on qubits, leading to problems such as the violation of the no-cloning theorem discussed in \cref{sec:qsharp}.
Qubits come into scope with either the \lqs{use} or \lqs{borrow} keywords.
The former provides access to freshly allocated qubits in state $\ket{0}$, while the latter allows access to previously allocated (and potentially entangled) qubits. We do not consider borrowing in this work as it is an optimization concern that lets a programmer
reuse ancillae in their code. The only allowed operations on qubits are gate application and measurement.

\subsection{Syntax}
\label{sec:syntax}

\begin{figure}[t]
  \small
\begin{synchart*}{syn}
  \Sort{Typ} & \tau & \bnfdef & \qrefty{q}             & \blue{\qrefty{q}}                          & \text{qubit reference} \\
             &      & \bnfalt & \arrty{\tau_1}{\tau_2} & \blue{\arrty*{\blk{\tau_1}}{\blk{\tau_2}}} & \text{function} \\
             &      & \bnfalt & \cmdty{\tau}           & \blue{\cmdty*{\blk{\tau}}}                 & \text{command} \\
             &      & \bnfalt & \vprodty{\tau}         & \blue{\vprodty*{\blk{\tau_i}}}             & \text{variadic product} \\
             &      & \bnfalt & \boolty                & \blue{\boolty}                             & \text{boolean} \\
             &      & \bnfalt & \unitty                & \blue{\unitty}                             & \text{unit} \\[1ex]
  \Sort{Exp} & e & \bnfdef & x                    & x                                              & \text{variable} \\
             &   & \bnfalt & \letbnd{e_1}{x}{e_2} & \blue{\letbnd*{\blk{e_1}}{x}{\blk{e_2}}}       & \text{let binding} \\
             &   & \bnfalt & \lamex{x}{e}         & \blue{\lamex*{x}{\blk{e}}}                     & \text{function} \\
             &   & \bnfalt & \appex{e_1}{e_2}     & \blue{\appex*{\blk{e_1}}{\blk{e_2}}}           & \text{application} \\
             &   & \bnfalt & \cmdex{m}            & \blue{\cmdex*{\blk{m}}}                        & \text{encapsulated command} \\
             &   & \bnfalt & \vtupleex{e}         & \blue{\vtupleex*{\blk{e}}}                     & \text{tuple} \\
             &   & \bnfalt & \vprojex{e}          & \blue{\vprojex*{\blk{e}}}                      & \text{projection} \\
             &   & \bnfalt & \trueex              & \blue{\trueex*}                                & \text{true} \\
             &   & \bnfalt & \falseex             & \blue{\falseex*}                               & \text{false} \\
             &   & \bnfalt & \ifex{e}{e_1}{e_2}   & \blue{\ifSugar{\blk{e}}{\blk{e_1}}{\blk{e_2}}} & \text{if expression} \\
             &   & \bnfalt & \unitex              & \blue{\unitex*}                                & \text{unit} \\[1ex]
  \Sort{Cmd} & m & \bnfdef & \retcmd{e}              & \blue{\retcmd*{\blk{e}}}                    & \text{return} \\
             &   & \bnfalt & \bndcmd{e}{x}{m}        & \blue{\bndcmd*{\blk{e}}{x}{\blk{m}}}        & \text{bind} \\
             &   & \bnfalt & \newcmd{x}{m}           & \blue{\newcmd*{x}{\blk{m}}}                 & \text{new qubit reference} \\
             &   & \bnfalt & \gateap{U}{e}           & \blue{\gateap*{U}{\blk{e}}}                 & \text{gate application} \\
             &   & \bnfalt & \diagap{U}{V}{e_1}{e_2} & \blue{\diagap*{U}{V}{\blk{e_1}}{\blk{e_2}}} & \text{diagonal gate application}   \\
             &   & \bnfalt & \meascmd{e}             & \blue{\meascmd*{\blk{e}}}                   & \text{measure}
\end{synchart*}
  \caption{Abstract and concrete syntax of \core.}
  \label{fig:abssyntax}
\end{figure}

\Cref{fig:abssyntax} presents the abstract syntax of \core. We divide the grammar into a monadic effectful command language and a pure expression language (the simply-typed $\lambda$-calculus extended with encapsulated commands).
We precisely specify the binding structure of the syntax following the notion of abstract binding trees from Harper's \PFPL~\autocite{Harper2016}.
Following the \PFPL-syntactic conventions, $\qrefty{q}$, $\gateap{U}{e}$, and $\diagap{U}{V}{e_1}{e_2}$ are indexed by symbols~\autocite[Ch. 31]{Harper2016} (marked in color) and variadic product operators are indexed by finite sets, $n$, where we slightly abuse the notation, $n \eqdef \{\,\kw{1},\kw{2},\dots,\kw{n}\,\}$. We also use the notation, $\tau_n \eqdef \finmap{n}{i}{\tau}$. Some operators take optional arguments marked by square brackets.
We will often use the concrete syntax in \blue{blue color}, and some standard derived forms from Harper's language \MAPFPL, shown in \cref{app:derived}, wherever there is no possibility of confusion.

The qubit reference type $\qrefty{\qbit{q}}$ is a singleton type~\autocite{Aspinall1995,Hayashi1994}, which is equivalent to \emph{ptr}$(l)$ in alias types~\autocite{Smith2000}.
Qubit symbols are shown in {\qbit{orange}} to distinguish them from the usual variables denoted by the metavariable $x$; we use qubit symbols to model the underlying logical qubit that the surface \qsharp language does not expose.
Unitary operations, $\igate{U}$ (shown in $\igate{pink}$), are parametric to the grammar; similar to \qsharp, which does not prefer a specific gate set. An $n$-qubit unitary is typed as $\igate{U} : \blue{ \funty*{\vprodty*<n>{\qrefty{q_\blk{i}}}}{\cmdty*{\unitty}}}$, where $\snsp{dim}(\igate{U}) = 2^n$. This type ensures that multi-qubit gates can be applied only to distinct qubits.
The $\blue{\gateap*{U}{e}}$ command applies the given unitary operation to a tuple of unique qubit references, where we follow singleton-tuple equivalence like \qsharp in case of a single-qubit unitary. Controlled unitaries can be represented using block diagonals, e.g., $\igate{CNOT} \triangleq \blue{\snsp{D}_{(\igate{I_{\blk{2}}},\igate{X})}}$ and are typed as $\blue{\snsp{D}_{(\igate{U},\igate{V})}} : \blue{ \funty*{\vprodty*<n+1>{\qrefty{q_\blk{i}}}}{\cmdty*{\unitty}}}$, where $\snsp{dim}(\igate{U}) = \snsp{dim}(\igate{V}) = 2^n$. It is understood that the number of arguments required for both forms of gate application depends on the dimension of the unitary parameters involved and is enforced by the typing rules.

\subsection{Static Semantics}
\label{sec:statics}

The pure fragment of \core is the usual simply-typed $\lambda$-calculus, so we will not say much about it here. We show typing rules for the effectful portion of \core in \Cref{fig:cmdtyping}.

\begin{figure}[t]
  \small
  \drules[cmd]{$\Gamma \vdash_\Sigma m \altdiv \tau$}{$m$ is a well-formed command relative to $\Sigma$, returning a value of type $\tau$}{Ret,Bnd,NewQRef,GateApRef,DiagApRef,MeasRef}
  \caption{Typing of commands. $\Gamma$ is the standard typing context, and the signature, $\Sigma$, keeps track of qubit symbols in scope. Each qubit symbol is required to be distinct. See other rules in \cref{app:statics}.}
  \label{fig:cmdtyping}
\end{figure}

All of our command typing judgments are parameterized by a signature, $\Sigma$, that keeps track of available qubit symbols in scope and corresponds to the shape of the quantum memory, much like store shapes\footnote{Store shapes follow laws similar to what are known as lenses in the current literature~\autocite{Foster2007}.} in the semantics of \algol~\autocite{Oles1982, Oles1985, Reynolds1997}.
The intuition behind incorporating a signature is that the block structure induced by the allocation command changes the shape of the quantum memory under consideration by making a new qubit available to the program on entry and removing it on exit. This is the essence of the stack-like treatment of local state.\footnote{Another view is to think of the commands as being parametrically polymorphic~\autocite{Fluet2006, OHearn2000, OHearn1997} to the store, an idea considered by Reynolds as early as 1975~\autocite{Brookes2014, Reynolds1975}, even before store shapes.
Still, we prefer the signature-based approach taken in Harper's \MAPFPL.}

\Rref{cmd-Ret,cmd-Bnd} are the two standard rules for monadic return and bind operations. The other four rules are specific to quantum computation.

The $\newAbtSugar{x}{m}$ command allocates a fresh logical qubit $\qbit{q}$ and immediately makes a reference to it available in the scope of $m$.
Its typing \rref{cmd-NewQRef} says that if command $m$ returns a value of type $\tau$ in a context containing $x : \qrefty{q}$ and a signature extended with $\qbit{q}$, then $\newAbtSugar{x}{m}$ returns a value of type $\tau$.
The binding structure ensures that the lifetime of the newly allocated qubit is equal to its lexical scope,
ensuring a strict stack discipline and providing safe and automatic management of qubits.

\Rref{cmd-GateApRef,cmd-DiagApRef} for unitary operations enforce the constraint that the input qubit references are distinct.
\Rref{cmd-MeasRef} shows how to obtain a boolean value from an expression that resolves to a qubit reference.

In summary, the allocation command changes the shape of the store (quantum state under consideration), while commands like unitary application and measurement change the store (quantum state).

\subsubsection{Safety Properties}
\label{para:safety}
We claim that our type system supports two safety properties currently not offered by \qsharp:
\begin{proposition}
  \core supports controlled aliasing and hence, statically enforces the no-cloning theorem for all unitary operations.
\end{proposition}

This follows from \rref{cmd-GateApRef,cmd-DiagApRef}: The premises of both typing rules require the input qubit references to be unique as all the entries in a tuple are required to be references to different logical qubits. In the case of the block diagonal, the control qubit reference, $e_1$, is also required to be distinct from the qubit references in $e_2$.

\begin{example}
  The unsafe code fragment from \cref{lst:alias} can be written in \core syntax as:
    \[ \newAbtSugar{q_1}{\retcmd*{\letAbtSugar{\blk{q_1}}{q_2}{\cmdex*{\diagap[2]{I}{X}{\blk{q_1}}{\blk{q_2}}}}}} \]
  or in the concrete syntax as
  \blue{$\newcmd*{q_1}{\retcmd*{\letbnd*{\blk{q_1}}{q_2}{\cmdex*{\diagap*{I_{\blk{2}}}{X}{\blk{q_1}}{\blk{q_2}}}}}}$}.
Since the type of the qubit reference in \core, $\qrefty{\qbit{q}}$, is indexed by the symbolic name of the qubit, we can tell statically that $q_1$ and $q_2$ reference the same underlying logical qubit. This allows our type system to reject the above program even though the \qsharp compiler allows it to pass.
\end{example}

\begin{proposition}
  \core statically ensures safe memory management and disallows dangling qubit references.
\end{proposition}

The allocation command, $\newAbtSugar{x}{m}$, as previously explained, comes with its own binding form, which ensures that the reference created during allocation can never escape its lexical scope. In \rref{cmd-NewQRef}, the fresh logical qubit $\qbit{q}$ allocated during this command is only available in the extended signature in the premise and not in the conclusion at the end of the command.

\begin{example}
  Using \core concrete syntax and the derived forms from \cref{app:derived}, the unsafe code fragment shown in \cref{lst:stack} can be written as
  \blue{$\letbnd*{\kw{proc}()~\{~\newcmd*{x}{\retcmd*{\blk{x}}}~\}}{\mathrm{NewQubit}}{\unitex*{}}$}.
  Here, while $\blue{\retcmd*{\blk{x}}}$ has type $\blue{\qrefty{q}}$ when $\qbit{q}$ is in the signature, i.e., $\Gamma, x: \blue{\qrefty{q}} \vdash_{\Sigma,\qbit{q}} \blue{\retcmd*{\blk{x}}} \altdiv \blue{\qrefty{q}}$; the conclusion of \rref{cmd-NewQRef} removes $\qbit{q}$ from scope and renders $\blue{\newcmd*{x}{\retcmd*{\blk{x}}}}$ ill-typed, i.e.,  $\Gamma \nvdash_{\Sigma} \blue{\newcmd*{x}{\retcmd*{\blk{x}}}} \altdiv \blue{\qrefty{q}}$.

\end{example}

\subsection{Dynamic Semantics}
\label{sec:dynamics}
As Staton~\autocite[p. 3]{Staton2015} suggests, ``by giving a fully complete equational theory we can understand quantum computation from the axioms of the theory without having to turn to denotational models built from operator algebra''; we rely on his equational theory for quantum local store~\autocite[\S 6.2]{Staton2015} to provide an equational dynamics for the effectful quantum fragment of \core.\footnote{We show the traditional operational semantics in \cref{app:transsem}.} Unlike Staton's language, our unitary operations do not return qubits but modify them in place. In this presentation, we use several derived forms from \Cref{app:derived}. Specifically, \blue{\kw{do}} returns the result of sequential execution of commands.
The program equations assume the availability of a universal gate set.

\paragraph{Interesting Axioms}
\begin{align*}
  a:\blue{\qrefty{q}} \vdash
  \blue{\docmd{\gateap*{X}{\blk{a}};\,\meascmd*{\blk{a}}}} \ &\equiv\ \blue{\neg~\docmd{\meascmd*{\blk{a}}}} \tag{A} \\
  \mathclap{a:\blue{\qrefty{q}},\ b:\blue{ \vprodty*<n>{\qrefty{r_\blk{i}}}} \vdash  \blue{\s{}\{\diagap*{U}{V}{\blk{a}}{\blk{b}};\, \meascmd*{\blk{a}};\, \retcmd*{\unitex*}\}} \equiv\qquad\qquad\qquad\quad\quad} \\
  \mathclap{\ \ \qquad\qquad \blue{\s{}\{\blk{x} \leftarrow \meascmd*{\blk{a}};\, \retcmd*{\ifSugar{\blk{x}}{\cmdex*{\gateap*{V}{\blk{b}}}}{\cmdex*{\gateap*{U}{\blk{b}}}}} \}}}  \tag{B} \\
  \cdot \vdash{}
  \blue{\docmd{\newcmd*{a}{\meascmd*{\blk{a}}}}} \ &\equiv\ \blue{\falseex} \tag{D} \\
  b:\blue{\qrefty{q}} \vdash{}
  \blue{\docmd{\newcmd*{a}{\diagap*{U}{V}{\blk{a}}{\blk{b}}}}} \ &\equiv\ \blue{\docmd{ \gateap*{U}{\blk{b}};\, \newcmd*{a}{\retcmd*{\unitex*}}}} \tag{E}
\end{align*}

As mentioned in \cref{sec:qramtheory}, we omit Staton's axiom (C) because, in the QRAM model, the discard operation just forgets the name of the reference to a qubit. In \qsharp and \core, we may consider an equivalent behavior: qubit references are automatically forgotten when they reach the end of their lexical scope.
A degenerate case of axiom (C) (for a $1 \times 1$ unitary) holds for both Staton's theory for a quantum local store and \core; it says that one can ignore the global phase.

\paragraph{Administrative Axioms}
The following equations correspond to respecting the composition and product monoidal structure of unitaries:
\begin{align}
  \mathclap{m_1:\blue{\cmdty*{\tau_1}},\ m_2:\blue{\cmdty*{\tau_2}} \vdash{}
  \blue{\docmd{ \newcmd*{a}{\newcmd*{b}{\blk{m_1}; \gateap*{SWAP}{\blk{a},\blk{b}};\, \blk{m_2}}}}} \ \equiv\qquad\qquad\qquad\qquad\qquad\qquad} \notag \\
  \mathclap{ \ \,\qquad \blue{\s{\kw{do}}\{\newcmd*{a}{\newcmd*{b}{\blk{m_1};\letbnd*{\pairex*{\blk{a}}{\blk{b}}}{\pairex*{\blk{b}}{\blk{a}}}{\cmdex*{\blk{m_2}}}}}}} \tag{F} \\
  e:\blue{ \vprodty*<n>{\qrefty{q_\blk{i}}}} \vdash{}\,
  \blue{\docmd{ \gateap*{I_{\blk{2^n}}}{\blk{e}}}} \ &\equiv\ \unitex*{} \tag{G} \\
  e:\blue{\vprodty*<n>{\qrefty{q_\blk{i}}}} \vdash{}
  \blue{\docmd{\gateap*{VU}{\blk{e}}}} \ &\equiv\ \blue{\docmd{\gateap*{U}{\blk{e}}; \gateap*{V}{\blk{e}}}} \tag{H} \\
  e_1:\blue{\vprodty*<m>{\qrefty{q_\blk{i}}}}, \qquad\qquad\qquad\qquad\qquad\ \notag \\
  e_2:\blue{\vprodty*<n>{\qrefty{r_\blk{i}}}} \vdash{}
  \blue{\docmd{ \gateap*{U\otimes V}{\blk{e_1,e_2}}}} \ &\equiv\ \blue{\docmd{ \gateap*{U}{\blk{e_1}}; \gateap*{V}{\blk{e_2}}}} \tag{I}
\end{align}

Selinger~\autocite{Selinger2004} notes that the \igate{SWAP} gate is equivalent to classically renaming qubit references, which captures the intuition behind equation (F). However, in our case, we have to ensure that the scope of the qubits is limited to our expression (since \igate{SWAP} is stateful). Axiom (G) says that applying an identity gate is equivalent to doing nothing. Axioms (H) and (I) show the two ways of composing unitaries---sequential and tensor products (horizontal and vertical composition, respectively, in circuit notation).

Like Staton, we also state the commutativity equations that hold for \core:
\begin{align}
  \mathclap{a:\blue{\qrefty{q}},\ b:\blue{\qrefty{r}},\ m: \blue{\cmdty{\blk{\tau}}} \vdash{} \blue{\docmd{ \bndSugar{\meascmd*{\blk{a}}}{x}{\bndSugar{\meascmd*{\blk{b}}}{y}{\blk{m}}}}} \ \equiv}\qquad\qquad \notag \\
  \mathclap{\ \ \quad\quad\qquad\qquad\blue{\docmd{ \bndSugar{\meascmd*{\blk{b}}}{y}{\bndSugar{\meascmd*{\blk{a}}}{x}{\blk{m}}}}}} \tag{J} \\
  m:\blue{\cmdty{\blk{\tau}}} \vdash{}
  \blue{\docmd{ \newcmd*{a}{\newcmd*{b}{\blk{m}}}}} \ &\equiv\ \blue{\docmd{ \newcmd*{b}{\newcmd*{a}{\blk{m}}}}} \tag{K} \\
  \mathclap{b:\blue{\qrefty{q}},\ m:\blue{\cmdty{\blk{\tau}}} \vdash{}
  \blue{\docmd{ \newcmd*{a}{\bndSugar{\meascmd*{\blk{b}}}{y}{\blk{m}}}}} \ \equiv} \notag \\
  \mathclap{\,\ \quad\qquad\qquad\qquad\qquad\blue{\docmd{ \bndSugar{\meascmd*{\blk{b}}}{y}{\newcmd*{a}{\blk{m}}}}}} \tag{L}
\end{align}

Now that we have shown that the quantum portion of \core is equivalent to Staton's quantum programming language~\autocite[\S 5]{Staton2015}\footnote{See \Cref{app:correspondence} for the trivial term translation.} and corresponding program equations with his theory of quantum local store, we can restate Staton's result~\autocite[p. 8, Theorem 11]{Staton2015} for our language:

\begin{theorem}[Universality of \core]
For any linear map $f : M_{2^{n_1}} \oplus \dots \oplus M_{2^{n_k}} \to M_{2^p}$ that is
completely positive and unital (i.e. corresponds to a trace-preserving superoperator), there is a $\core$ term, $t$, such that $t$ implements $f$.
\end{theorem}

This corresponds to Staton's Theorem 11.1~\autocite{Staton2015}, which is a variation on Selinger's Theorem 6.14~\autocite{Selinger2004}. The proof relies on the correspondence between Staton's simple quantum language and a fragment of \core, where the translation is straightforward (\Cref{app:correspondence}).

\begin{theorem}[Completeness]
    Assuming an axiomatization of unitaries, if two terms $t$ and $u$ have equivalent interpretation in a common context, $\Gamma$, then $\Gamma \vdash t \equiv u$ is derivable.
\end{theorem}

Note that since \qsharp is parameterized over gate sets, we need an equational theory over unitaries for \core. In the simplest case, we can declare two unitaries equal if their corresponding matrices are equal. Again, the result follows from Staton's Theorem 11.2. There are two differences: (1) instead of algebraic operations, our theorem is stated in terms of generic effects (which correspond directly to programming); (2) in addition to the equations (A)--(L) stated above, we also need the standard $\beta\eta$-equalities of simply-typed $\lambda$-calculus, which are required because our axioms do not live in isolation but are written as typed expressions in a context.

\section{Translation from \qsharp to \texorpdfstring{$\bm{\lambda}_{\mathbf{\qsharp}}$}{λ-\qsharp}}
\label{sec:elaboration}

\begin{table}[t]
    \centering
    \begin{tabular}{ll}
        \qsharp Syntax & \core Translation \\
        \midrule
        $\elab{(\tau_1,\dots,\tau_n)}$ & $\blue{ \vprodty*<n>{\elab{\tau_i}}}$ \\
        $\elabl\tau_1$~~\lqs{->} $\tau_2\elabr$ & $\elab{\tau_1}~ \blue{\to}~ \elab{\tau_2}$ \\
        $\elabl\tau_1$~\lqs{=>} $\tau_2\elabr$ & $\elab{\tau_1}~ \blue{\Rightarrow}~ \elab{\tau_2}$ \\
        $\elabl$\lqsnsp{Bool}$\elabr$ and $\elabl$\lqsnsp{Result}$\elabr$ & \blue{$\boolty$} \\
        $\elabl$\lqsnsp{Qubit}$\elabr$ & \blue{$\qrefty{q}$} \\
        $\elabl$\lqsnsp{Unit}$\elabr$ & \blue{$\unitty$} \\
        $\elabl$\lqs{function} $f$ ($x_1 : \tau_1$,~\dots) : $\tau$ \{$e$\}$\elabr$ &
         \blue{ $\lamex[\vprodty*<n>{\elab{\tau_i}}]{\vtupleex*<n>{x}}[\qquad\,\;\tau]{\elab{e}}$ } \\
        $\elabl$\lqs{operation} $f$ ($x_1 : \tau_1$,~\dots) : $\tau$ \{$s$\}$\elabr$ &
         \blue{ $\lamex[\vprodty*<n>{\elab{\tau_i}}]{\vtupleex*<n>{x}}[\cmdty{\tau}]{\elab{s}}$ } \\
        $\elabl$\lqs{return} $e\elabr$ & \blue{ $\retcmd*{\elab{e}}$ }  \\
        $\elabl$\lqs{let} $x$ = $e$;~\dots$\elabr$ & \blue{ $\letbnd*{\elab{e}}{x}{\elab{\dots}}$} \\
        $\elabl$\lqs{if} $e$ \{ $s_1$ \} \lqs{else} \{ $s_2$ \}$\elabr$ &
         \blue{ $\ifSugar{\elab{e}}{\cmdex*{\elab{s_1}}}{\cmdex*{\elab{s_2}}}$} \\
        $\elabl$\lqs{use} $q$ = \lqs{Qubit} \{ $s$ \}$\elabr$ & \blue{$\newcmd*{q}{\elab{s}}$} \\
        $\elabl$\lqs{Adjoint} $e_1$ ($e_2$)$\elabr$ & $\blue{ \gateap*{U^\dagger}{\blk{\elab{e_2}}}}$, where $\igate{U}=\snsp{mat}(\elab{e_1})$ \\
        $\elabl$\lqs{Controlled} $e_1$ ($q$, $e_2$)$\elabr$ & $\blue{ \diagap*{I_{\blk{2^n}}}{U}{\blk{q}}{\elab{e_2}}}$, where $\igate{U}=\snsp{mat}(\elab{e_1})$ \\
        $\elab{e_1~ (e_2)}$ & $\elab{e_1}\blue{(}\elab{e_2}\blue{)}$ \\
        $\elab{(e_1,\dots,e_n)}$ & $\blue{\langle} \elab{e_i},\dots,\elab{e_n} \blue{\rangle}$ \\
        $\elabl$\lqsnsp{true}$\elabr$ and $\elabl$\lqsnsp{One}$\elabr$ & \blue{$\trueex$} \\
        $\elabl$\lqsnsp{false}$\elabr$ and $\elabl$\lqsnsp{Zero}$\elabr$ & \blue{$\falseex$} \\
    \end{tabular}
    \caption{Select \qsharp to \core elaboration rules. $f$, $x$, and $q$ are variable names, $e$ is a \qsharp expression, $s$ is a \qsharp statement, and $\tau$ is a \qsharp type. $\elab{\cdot}$ is the elaboration function and mat$(\cdot)$ converts a \core expression to its corresponding unitary operator. In the rule for \lqs{Qubit}, $\qbit{q}$ is determined from the elaboration context.}
    \label{tab:elab}
\end{table}

We summarize the rules for converting from the supported features of \qsharp to \core in \Cref{tab:elab}.
For ease of presentation, we use the derived forms from \Cref{app:derived}.
\Cref{fig:elab-example} in \Cref{app:elaboration} shows the elaboration of the \qsharp teleport example from \Cref{lst:teleport}.

Elaboration maintains a \emph{context} (not shown in \Cref{tab:elab}) that stores the logical qubit associated with each qubit reference.
To translate the \qsharp type \lqs{Qubit} to the \core type  \blue{$\qrefty{q}$}, we look up the reference associated with the \lqs{Qubit} type in the context or add a new logical qubit to the context.
The \lqs{use} statements and \lqs{operation} parameters update the context to include a mapping from the new qubit or qubit parameter(s) to a fresh logical qubit.
In \Cref{fig:elab-example}, $\qbit{a}$, $\qbit{b}$, and $\qbit{m}$ are distinct logical qubits introduced by elaboration.

Elaboration performs some type checking to produce well-formed \core terms.
For example, elaboration checks that the first argument to a \lqs{Controlled} or \lqs{Adjoint} functor is equipped with the \lqs{Ctl} and/or \lqs{Adj} characteristics and inlines the corresponding operation specialization, converting it to a unitary operator.
We need to do this during elaboration because we do not yet encode characteristic information or specializations in \core.
An adjointable \qsharp operation with type \lqs{(Qubit,...,Qubit) => Unit} can be converted into a unitary matrix by composing the unitary representations of its primitive gates. Note that the type signature and the fact that the operation is adjointable (i.e., no measurement) mean it can be unfolded to a sequence of primitive gates.
We also expand multi-controlled operations (\lqs{Controlled} statements with a list of controls) into a nested group of single-qubit controlled operations, and expand \lqs{if-elif-else} expressions into nested \blue{if-then-else} expressions using \blue{$\langle\rangle$} in place of an empty \lqs{else} block.
Finally, we restrict \qsharp \lqs{function} bodies to be pure expressions since we do not handle classical \lqs{mutable} values.

As this is our initial attempt to get the foundations right, we do not yet support several \qsharp features: namespaces; operation characteristics; custom operation specializations (i.e., implementations of controlled or adjoint variants); general application of the \lqs{Adjoint} and \lqs{Controlled} functors; arrays and slices; type parameters; base types outside of \lqs{Bool}, \lqs{Result}, and \lqs{Unit}; iteration using \lqs{for}, \lqs{while}, or \lqs{repeat}; and \lqs{within-apply} blocks (which apply an operation and its adjoint). We say more in \cref{sec:conclusion} about the challenges involved in supporting some of these features.

\section{Related Work}
\label{sec:relatedwork}

\paragraph{Large Language Definition Efforts}

In starting this project, we were encouraged by previous efforts in the formal specification of large programming languages such as Standard ML~\autocite{Harper2000,Lee2007}, Java~\autocite{Igarashi2001}, JavaScript~\autocite{Guha2010}, Rust~\autocite{Jung2020,Jung2017}, and, most recently, Go~\autocite{Griesemer2020}. As we mentioned in \cref{sec:lambdaqs}, we more or less followed the pioneering methodology of the formalization and mechanization of the definition of Standard ML~\autocite{Milner1997} by identifying a well-founded core language and performing all metatheoretical reasoning on that core. These projects demonstrate the extent to which it is possible to distill large and complex languages into their formal and faithful essence. Java and Go serve as examples of industry-scale languages in mass use benefiting from formalization and academic study: Extensions such as generics (polymorphism) were first investigated on smaller cores of the respective languages before being adopted in production over the years. In the case of JavaScript, perhaps the impact of a careful formal study was even more significant as JavaScript is the de-facto programming language of the web. We hope our work serves as a similar playground for extensions and future impact.

\paragraph{Equational Theories}

Like Staton, we do not focus on the axiomatization of unitaries but of quantum computation in general. We discuss two similar works here.

Paykin and Zdancewic~\autocite{Paykin2019} build upon Staton's work and present an equational theory for quantum computation embedded inside homotopy type theory (HoTT)~\autocite{Univalent2013}. The essential idea to treat unitaries as higher inductive paths simplifies the presentation of the equational theory as several axioms can be derived using the rich structure of HoTT. While their work focuses on embedding a quantum language inside a highly expressive dependent type theory, we are motivated by practical concerns in defining semantics for a real-world quantum language.

Peng et al.~\autocite{Peng2022} introduce Non-Idempotent Kleene Algebra (NKAT) to reason about programs algebraically. Their language is based on Kozen's Kleene Algebra with Test (or KAT), which models both programs and assertions, allowing for a lightweight implementation of a Hoare-style logic~\autocite{Kozen1997}. While the underlying language of regular expressions is not designed for convenience in programming, their use of NKAT to verify quantum program transformations is a key use case of equational theories and one we plan to explore in the future.

\paragraph{Linearity and Monadic Quantum Languages}

Research-oriented languages like \qwire~\autocite{Paykin2017} and Silq~\autocite{Bichsel2020} employ a linear type system to enforce the no-cloning theorem. So far, industry languages, including \qsharp, have not adopted linear typing. The lack of linear typing in \qsharp is justified by its monadic treatment of state. That is, the monad interface imposes a sequential order to manipulate the quantum state as every monad can be treated as a linear-use state monad~\autocite{Mogelberg2014}. The design decision in \qsharp to permit uncontrolled aliasing of qubits for user comfort is the only reason a monadic interface is not enough, which is addressed by our type system.

Other monadic languages include Quantum IO Monad~\autocite{Altenkirch2009} (QIO) and Quantum Hoare Type Theory~\autocite{Singhal2020, Singhal2021} (QHTT). QIO is a pioneering monadic interface that isolates quantum effects inside a monad as we do in \core. QHTT is a typed framework that extends the QIO monad with pre- and postconditions so that precise specifications about the quantum state can be stated and proved in a dependent type theory.

\paragraph{\ALGOL-like Quantum Languages}
The IQu language~\autocite{Paolini2019} extends Idealized \algol with quantum circuits and quantum variables, much as we extend Harper's Modernized Algol (\MAPFPL). Like \core, IQu uses references to access qubits and therefore does not need a linear type system to prevent cloning. Though every newly allocated qubit is unique, IQu does not have a way to guarantee that multiple references to the same qubit are not passed to a single operation. Instead, IQu's use of Idealized \algol is focused on programmability, following a design philosophy similar to that of \qsharp. IQu allows programmers to write the classical parts of their programs in a familiar way while providing access to the quantum state.

\section{Conclusion and Perspectives}
\label{sec:conclusion}
We present a core calculus for the \qsharp programming language, dubbed \core. We maintain a separation between the quantum effectful and the pure expression sub-languages, expose the monadic nature of computation inherent in \qsharp, make qubit aliasing and block structure explicit, and present an equational semantics for \core, building upon Staton's fully complete equational theory for quantum computation.

A formal specification of the whole \qsharp language still requires more work. Some extensions are straightforward; e.g., classical mutable bindings in \qsharp can be modeled after assignables in Harper's \MAPFPL; conveniently, they follow the model of classical local store analogous to how we modeled the quantum local store in this paper. Here it helps that \qsharp does not allow references to any types other than qubits.
Other features are more challenging, including arrays, slices, iteration, polymorphism, and patterns like \lqs{within-apply} and repeat-until-success~\autocite{Paetznick2014}. Then there is the question of how to treat operations that have specializations supplied by the programmer versus those auto-generated by the \qsharp compiler (which is not known statically); we may need to consider a phase distinction~\autocite{Harper2021} here to distinguish between what can be derived statically using types and what requires inspecting the code.

We plan to gain confidence in our formalization by mechanizing its metatheory.
We see potential in recent developments such as the Agda-based formalization of Second-Order Abstract Syntax~\autocite{Fiore2022}, which lets users concisely specify algebraic theories such as Staton's and significantly reduces the boilerplate code required to state interesting theorems about the theory.
However, this tool does not support substructural assumptions on qubit symbols, making our proposed extension a nontrivial prospect.

A major goal of this project is to form a playground for prototyping extensions to the \qsharp type system.
For instance, a peculiar decision in \qsharp is to allow uncontrolled aliasing of qubits to support user-friendly features such as qubit arrays.
While convenient, reasoning about interference freedom for arrays is notoriously hard; specifically, our approach to enforce no-cloning inspired by alias types~\autocite{Smith2000, Walker2001a} does not easily scale to arrays~\autocite[\S3.5.1]{Walker2001}.
We are extending our \core type checker with a constraint solver to evaluate potential solutions for scenarios that occur in practice in \qsharp library code.
Depending on the complexity of the array indexing used in practice, we may use a natural number inequality checker, a simple symbolic numerical solver, or a full-fledged SMT solver like Z3~\autocite{deMoura2008} to guarantee qubit distinctness.

We could also statically check \qsharp's \lqs{Adj} and \lqs{Ctl} characteristics for validity. In the simplest case, we would flag operations as unitary or non-unitary in order to inform the compiler when adjoints and controls can be trivially synthesized, in the manner of Silq~\autocite{Bichsel2020}. However, more complex programs are adjointable and controllable in practice, which may require a more sophisticated approach.

One of our insights from this project is that even though quantum computation is a fundamentally new abstraction, many classical techniques from programming languages and compilers communities can be adapted to the quantum setting~\autocite{Svore2006}. \qsharp and its Quantum Development Kit (QDK) are significant examples of realizing that vision~\autocite{Aho2022}.
As a high-level programming language, \qsharp must also compile to efficient, low-level machine instructions. Recently, Microsoft announced QIR, a Quantum Intermediate Representation based on the popular LLVM framework~\autocite{Geller2020a}, which has gained significant industry backing in the form of the QIR Alliance~\autocite{QIRSpec2021}. This provides an exciting avenue for future development. We plan to explore semantics-preserving compilation from \qsharp to QIR using our formalization. This project will require formally specifying the semantics of QIR, for which we will draw upon the Verified LLVM (Vellvm) project~\autocite{Zakowski2021, Zhao2012}. We also aim to formalize QIR's \emph{profiles}, which specify what kinds of quantum operations are allowed on a given quantum architecture. This, along with our current work, will constitute a significant step toward our broader vision of a fully verified quantum stack~\autocite{Rand2019}.

\section*{Acknowledgments}
\phantomsection\addcontentsline{toc}{section}{Acknowledgments}
We thank Bettina Heim and Alan Geller for helping us get this project off the ground, Bob Harper for several fruitful conversations, Mike Hicks for pointing us to Alias Types during a preliminary presentation at PLanQC 2021, and Ohad Kammar for making several valuable connections. Jennifer Paykin, Sam Staton, Bob Harper, and anonymous reviewers from the PC of FSCD 2022 gave critical feedback on a previous draft. For presenting \core syntax, Bob Harper's PFPL syntax macros\footnote{Available on GitHub at \url{https://github.com/RobertHarper/pfpl-syntax}.} were very helpful. We also thank Matt Amy, Adrian Lehmann, and the anonymous reviewers of QPL 2022 for their feedback on the manuscript.

\paragraph{Funding}
This material is supported by EPiQC, an NSF Expedition in Computing, under Grant No. CCF-1730449, the Air Force Office of Scientific Research under Grant No. FA95502110051 and the U.S. Department of Energy, Office of Science, Office of Advanced Scientific Computing Research, Quantum Testbed Pathfinder Program under Award Number DE-SC0019040.

\phantomsection
\printbibliography[heading=bibintoc]

\newpage

\appendix

\section{Derived Forms}
\label{app:derived}

In addition to the syntax shown in \Cref{fig:abssyntax}, we use these straightforward derived forms from Harper's language \MAPFPL~\autocite{Harper2016a}:

\begin{align*}
  \blue{\{\blk{x} \leftarrow \blk{m_1}; \blk{m_2}\}} &\quad\triangleq\quad \blue{\bndcmd*{\cmdex*{\blk{m_1}}}{x}{\blk{m_2}}} \\
  \blue{\{\blk{x_1} \leftarrow \blk{m_1}; \dots \blk{x_{n-1}} \leftarrow \blk{m_{n-1}}; \blk{m_n}\}} &\quad\triangleq\quad \blue{\{\blk{x_1} \leftarrow \blk{m_1}; \dots \{\blk{x_{n-1}} \leftarrow \blk{m_{n-1}};  \blk{m_n}\}\}} \\
  \blue{\{\blk{m_1}; \blk{m_2}\}} &\quad\triangleq\quad \blue{\{ \_ \leftarrow \blk{m_1}; \blk{m_2}\}} \\
  \blue{\{\blk{m_1}; \dots \blk{m_{n-1}}; \blk{m_n}\}} &\quad\triangleq\quad \blue{\{\blk{m_1}; \dots \{\blk{m_{n-1}}; \blk{m_n}\}\}} \\
  \blue{\s{\kw{do}} \blk{m}} &\quad\triangleq\quad \blue{\{\blk{x} \leftarrow \blk{m}; \retcmd*{\blk{x}}\}} \\
  \blue{\blk{\tau_1} \Rightarrow \blk{\tau_2}} &\quad\triangleq\quad \blue{\blk{\tau_1} \rightarrow \cmdty{\blk{\tau_2}}} \\
  \blue{\s{\kw{proc}}(\blk{x:\tau})~\s{\blk{m}}} &\quad\triangleq\quad \blue{ \lamex*[\tau]{x}[\tau']{\cmdex*[\tau']{\blk{m}}}} \\
  \blue{\s{\kw{call}} \blk{e_1}(\blk{e_2})} &\quad\triangleq\quad \blue{\s{\kw{do}}(\appex*{\blk{e_1}}{\blk{e_2}})} \\
  \blue{\s{\kw{call}} \blk{e}} &\quad\triangleq\quad \blue{\s{\kw{call}} \blk{e}(\blk{\unitex*{}})} \\
\end{align*}

\section{Remaining Static and Dynamic Rules}
\label{app:rules}

These are standard rules from Harper's \PFPL~\autocite{Harper2016} adapted to quantum computation. Note that in the \PFPL terminology, we are following the \textit{scoped dynamics} of symbols~\autocite[Ch. 31]{Harper2016}. Instead of typed assignables~\autocite[\S 34.3]{Harper2016}, we only have a single type of qubit symbols, which we hence do not annotate in the signature, $\Sigma$, i.e., the signature only contains active qubit symbols in scope and nothing else. Further, since there are no reference types~\autocite[Ch. 35]{Harper2016} except for a single qubit reference type, we do not explicitly state any mobility conditions~\autocite[\S31.1]{Harper2016}. Under scoped dynamics, qubit references are immobile~\autocite[\S35.2]{Harper2016}. This mobility restriction is crucial in ensuring the stack discipline for qubit management.

\subsection{Type System}
\label{app:statics}
We provide the most interesting rules of our type system in \cref{fig:cmdtyping}. We include the following rules here for completeness.

\drules[ty]{$\Gamma \vdash e : \tau$}{Expression $e$ has type $\tau$ in context $\Gamma$}{Var,Let,Lam,Ap,Pr,Tpl}

In \rref{tyS-QLoc} and \rref{vS-QLoc} in the next subsection, $\mathbf{qloc}[\qbit{q}]$ is the value of the reference to an active qubit symbol $\qbit{q}$. It can be thought of as a classical pointer value indexed by a qubit symbol. The signature plays a role wherever commands or qubits are involved.

\drules[tyS]{$\Gamma \vdash_\Sigma e : \tau$}{Expression $e$ has type $\tau$ relative to the signature}{Cmd,QLoc}

\subsection{Operations Semantics}
\label{app:transsem}

\paragraph{Pure Classical Sub-language}

\drules[v]{$e$ \kw{val}}{$e$ is a value}{Lam,Tpl}

\drules[tr]{$e \longmapsto e'$}{$e$ steps to $e'$}{Let,LetInstr,ApL,ApR,ApInstr,Tpl,Pr,PrInstr}

\drules[vS]{$e$ \kw{val}$_\Sigma$}{$e$ is a value relative to $\Sigma$}{Cmd,QLoc}

\paragraph{Effectful Quantum Sublanguage}

In the following rules, we do not show the quantum store, preferring the equational dynamics shown in \cref{sec:dynamics}. In reading these rules, consider the quantum store shape being expanded and restored during the allocation command (as reflected in the signature) and the quantum state being modified during the measurement and the gate application commands.

\drules[fn]{$m$ \kw{final}$_\Sigma$}{Command $m$ is complete}{Ret}

\drules[st]{$m \longmapsto_\Sigma m'$}{Command $m$ steps to $m'$}{Ret,Bnd,BndInstr,BndCmd,NewQRef,NewQRefInstr,GateApRef,DiagApRefL,DiagApRefR,MeasRef}

\section{Correspondence between \texorpdfstring{$\bm{\lambda}_{\mathbf{\qsharp}}$}{λ-\qsharp} and Staton's quantum language}
\label{app:correspondence}

We can easily translate the quantum-specific fragment of \core to Staton's quantum programming language~\autocite[\S 5]{Staton2015}. Note that the generic effects of his quantum language are equivalent to the algebraic operations (of the algebraic theory)~\autocite{Plotkin2003} that he uses in his proof of Theorem 11~\autocite[pp. 12--15, Appendix A]{Staton2015}. The translation follows:

\begin{align*}
  \s{let} q = \su{new}()~\s{in} m &\ \equiv\ \newAbtSugar{q}{m} \\
 \su{measure}(e)                  &\ \equiv\ \meascmd{e} \\
 \su{apply}_U(e)                  &\ \equiv\ \gateap{U}{e}
\end{align*}

Note that $\gateap{U}{\blk{e}}$ subsumes $\diagap{U}{V}{e_1}{e_2}$, just like in Staton's work. In other words, we can define each of Staton's terms using terms of our language.

\section{An Elaboration Example}
\label{app:elaboration}

\Cref{lst:teleport} shows a sample \qsharp program. \Cref{fig:elab-example} shows the corresponding elaboration to \core.

\begin{lstlisting}[label=lst:teleport,float=ht,captionpos=b,caption={Teleportation in \qsharp (adapted from Quantum Katas~\autocite{Mykhailova2020}).}]
namespace Quantum.Kata.Teleportation {

    open Microsoft.Quantum.Intrinsic; // for H, X, Z, CNOT, and M

    operation Entangle (qAlice : Qubit, qBob : Qubit) : Unit is Adj {
        H(qAlice);
        CNOT(qAlice, qBob);
    }

    operation SendMsg (qAlice : Qubit, qMsg : Qubit) : (Bool, Bool) {
        CNOT(qMsg, qAlice);
        H(qMsg);
        return (M(qMsg) == One, M(qAlice) == One);
    }

    operation DecodeMsg (qBob : Qubit, (b1 : Bool, b2 : Bool)) : Unit {
        if b1 { Z(qBob); }
        if b2 { X(qBob); }
    }

    operation Teleport (qAlice : Qubit, qBob : Qubit, qMsg : Qubit) : Unit {
        Entangle(qAlice, qBob);
        let classicalBits = SendMsg(qAlice, qMsg);
        DecodeMsg(qBob, classicalBits);
    }
}
\end{lstlisting}

\begin{figure}
  \blue{\begin{align*}
  \s{\kw{let}}&\blk{\s{Entangle}}\s{\kw{be~proc}}( \langle \blk{\snsp{qAlice}}, \blk{\snsp{qBob}} \rangle : \qrefty{a} \times \qrefty{b} ) ~ \{ \\
  &\gateap*{H}{\blk{\snsp{qAlice}}}; \\
  &\diagap*{I_2}{X}{\blk{\snsp{qAlice}}}{\blk{\snsp{qBob}}}\}~\s{\kw{in}} \\
  \s{\kw{let}}&\blk{\s{SendMsg}} \s{\kw{be~proc}}( \langle \blk{\snsp{qAlice}}, \blk{\snsp{qMsg}} \rangle : \qrefty{a} \times \qrefty{m} ) ~\{ \\
  &\diagap*{I_2}{X}{\blk{\snsp{qMsg}}}{\blk{\snsp{qAlice}}}; \\
  &\gateap*{H}{\blk{\snsp{qMsg}}}; \\
  &\retcmd*{\langle \cmdex*{\meascmd*{\blk{\snsp{qMsg}}}}, \cmdex*{\meascmd*{\blk{\snsp{qAlice}}}} \rangle}\}~\s{\kw{in}} \\
  \s{\kw{let}}&\blk{\s{DecodeMsg}} \s{\kw{be~proc}}( \langle \blk{\snsp{qBob}}, \langle \blk{\snsp{b_1}}, \blk{\snsp{b_2}} \rangle \rangle : \qrefty{b} \times (\snsp{bool} \times \snsp{bool})) ~\{ \\
  & \ifSugar{\blk{\snsp{b_1}}}{\gateap*{Z}{\blk{\snsp{qBob}}}}{\unitex*{}}; \\
  & \ifSugar{\blk{\snsp{b_2}}}{\gateap*{X}{\blk{\snsp{qBob}}}}{\unitex*{}}\}~\s{\kw{in}} \\
  \s{\kw{let}}&\blk{\s{Teleport}} \s{\kw{be~proc}}( \langle \blk{\snsp{qAlice}}, \blk{\snsp{qBob}}, \blk{\snsp{qMsg}} \rangle : \qrefty{a} \times \qrefty{b} \times \qrefty{m} ) ~\{ \\
  &\s{\kw{call}}\blk{\snsp{Entangle}} (\langle \blk{\snsp{qAlice}}, \blk{\snsp{qBob}} \rangle); \\
  &\blk{\s{classicalBits}} \leftarrow \s{\kw{call}}\blk{\snsp{SendMsg}}(\langle \blk{\snsp{qAlice}}, \blk{\snsp{qMsg}} \rangle); \\
  &\s{\kw{call}}\blk{\snsp{DecodeMsg}} (\langle \blk{\snsp{qBob}}, \blk{\snsp{classicalBits}} \rangle)\}~\s{\kw{in}} \unitex*{}
  \end{align*}}
    \caption{\core elaboration of the \qsharp program in \Cref{lst:teleport}.}
    \label{fig:elab-example}
  \end{figure}

\end{document}